\documentclass[aps,prl,reprint,preprintnumbers,showpacs,floatfix,nofootinbib,superscript address,longbibliography]{revtex4-1}
\usepackage[utf8]{inputenc}
\usepackage{amssymb}
\usepackage{hhline}
\usepackage{amsmath}
\usepackage{mathtools}
\usepackage[dvipsnames]{xcolor}
\usepackage{multirow,tabularx}
\usepackage{graphicx}
\usepackage{natbib}

\usepackage{xspace}
\usepackage{xstring}
\usepackage{titlesec}
\usepackage{parskip}
\allowdisplaybreaks
\usepackage{braket}

\newcommand*\diff{\mathrm{d}}
\newrobustcmd{\pea}[1]{%
	\emph{#1}\textbf{\ \  ---\ }
}
\titleformat{\paragraph}[runin]{\normalfont\normalsize\bfseries}{\emph\theparagraph}{1em}{\pea}

\newcommand*{\ie}{i.e.\@\xspace}
\newcommand*{\eg}{e.g.\@\xspace}
\newcommand*{\cf}{c.f.\@\xspace}
\newcommand*{\fig}{fig.\@\xspace}
\newcommand*{\figs}{figs.\@\xspace}
\newcommand*{\eq}{eq.\@\xspace}
\newcommand*{\eqs}{eqs.\@\xspace}
\newcommand*{\wrt}{w.r.t.\@\xspace}

\renewcommand*{\S}{S}

\usepackage{hyperref}
\hypersetup{%
	colorlinks = true,%
	linkcolor = Blue,%
	citecolor = Blue,%
	filecolor = Blue,%
	urlcolor = Blue%
}%

\begin{document}

\title{Transitioning to Memory Burden:\\
Detectable Small Primordial Black Holes as Dark Matter}

\author{Gia Dvali}
\email{gdvali@mpp.mpg.de}
\affiliation{Arnold Sommerfeld Center, Ludwig-Maximilians-Universit{\"a}t, Theresienstr.~37, 80333 M{\"u}nchen, Germany}
\affiliation{Max-Planck-Institut f{\"u}r Physik, Boltzmannstr.~8, 85748 Garching, Germany}

\author{Michael Zantedeschi}
\email{michael.zantedeschi@pi.infn.it}
\affiliation{INFN, Sezione di Pisa,
Largo Bruno Pontecorvo 3, I-56127 Pisa, Italy}

\author{Sebastian Zell}
\email{sebastian.zell@lmu.de}
\affiliation{Arnold Sommerfeld Center, Ludwig-Maximilians-Universit{\"a}t, Theresienstr.~37, 80333 M{\"u}nchen, Germany}
\affiliation{Max-Planck-Institut f{\"u}r Physik, Boltzmannstr.~8, 85748 Garching, Germany}

\begin{abstract}
     Mounting theoretical evidence suggests that black holes are subjected to 
    the memory burden effect, implying that after certain time the 
    information stored in them suppresses the decay rate. 
    This effect opens up a new window for small primordial black holes (PBHs) below $10^{15}\,{\rm g}$ as dark matter. 
    We show that the smooth transition from semi-classical evaporation to the memory-burdened phase strongly impacts observational bounds on the abundance of small PBHs.  The most stringent constraints come from present-day fluxes of astrophysical particles.
    Remarkably, currently-transitioning small PBHs are detectable through high-energetic neutrino events. 
\end{abstract}

\maketitle

\paragraph*{Introduction} The microscopic origin of dark matter
is among the greatest mysteries of nature. 
Primordial black holes (PBHs) formed in the early Universe~\cite{Zeldovich:1967lct, Hawking:1971ei,Carr:1974nx,Chapline:1975ojl,Carr:1975qj} stand out as 
highly-motivated dark matter candidate (for recent reviews, see \eg~\cite{Carr:2020xqk,Green:2020jor}). One of the most important observational constraints on PBHs comes from Hawking's famous demonstration~\cite{Hawking:1975vcx} that they  
radiate and thereby gradually lose mass.  Thus, only the PBHs that have survived until today can potentially account for  dark matter.  

In the semi-classical approximation, a black hole of mass $M$  loses mass at the rate~\cite{Hawking:1975vcx} $\left(\diff M(t)/\diff t\right)_{\text{sc}}  \simeq  r_g^{-2} $,
where $r_g = 2 GM$ is the Schwarzschild radius and $G$ denotes Newton's constant. If this were applicable throughout the entire lifetime, a black hole of initial mass $M_0$ would evaporate completely in a time given by
\begin{equation} \label{tSC}
	 \tau_{\text{sc}} \simeq M_0\left(\frac{\diff M(t)}{\diff t}\right)_{\text{sc}} ^{-1} \simeq \S\, r_g \;.
\end{equation}
We expressed this result in terms of the Bekenstein-Hawking entropy~\cite{Bekenstein:1973ur} 
\begin{equation} \label{entropy}
	\S = 2\pi M r_g \simeq 10^{30} \times \!\left(\frac{M}{10^{10}\,{\rm g}}\right)^{\!2}\;,
\end{equation} 
which in natural units ($\hbar = c =1$) is the unique dimensionless characteristic of a black hole. If \eq \eqref{tSC} were to hold, black holes lighter than $\sim 10^{15}\,{\rm g}$ would evaporate in less than the age of the Universe.

However, the standard constraints on PBHs~\cite{Carr:2009jm,Carr:2016hva,Carr:2020gox} have been derived under the assumption of validity of Hawking's 
  semi-classical regime during the 
  entire period of black hole evaporation. 
  This assumption does not take into account the 
  quantum back reaction on the evaporating black hole,  
  manifesting in the so-called ``memory burden" (MB) effect~\cite{Dvali:2018xpy,Dvali:2018ytn,Dvali:2020wft,Alexandre:2024nuo, Dvali:2024hsb}.
  This effect represents a generic phenomenon 
  exhibited by the systems of high capacity of information storage. Its essence is that the information stored by the 
  system back reacts and tends to stabilize it against the decay. 

\begin{figure}
\centering
    \includegraphics[width=1.\linewidth]{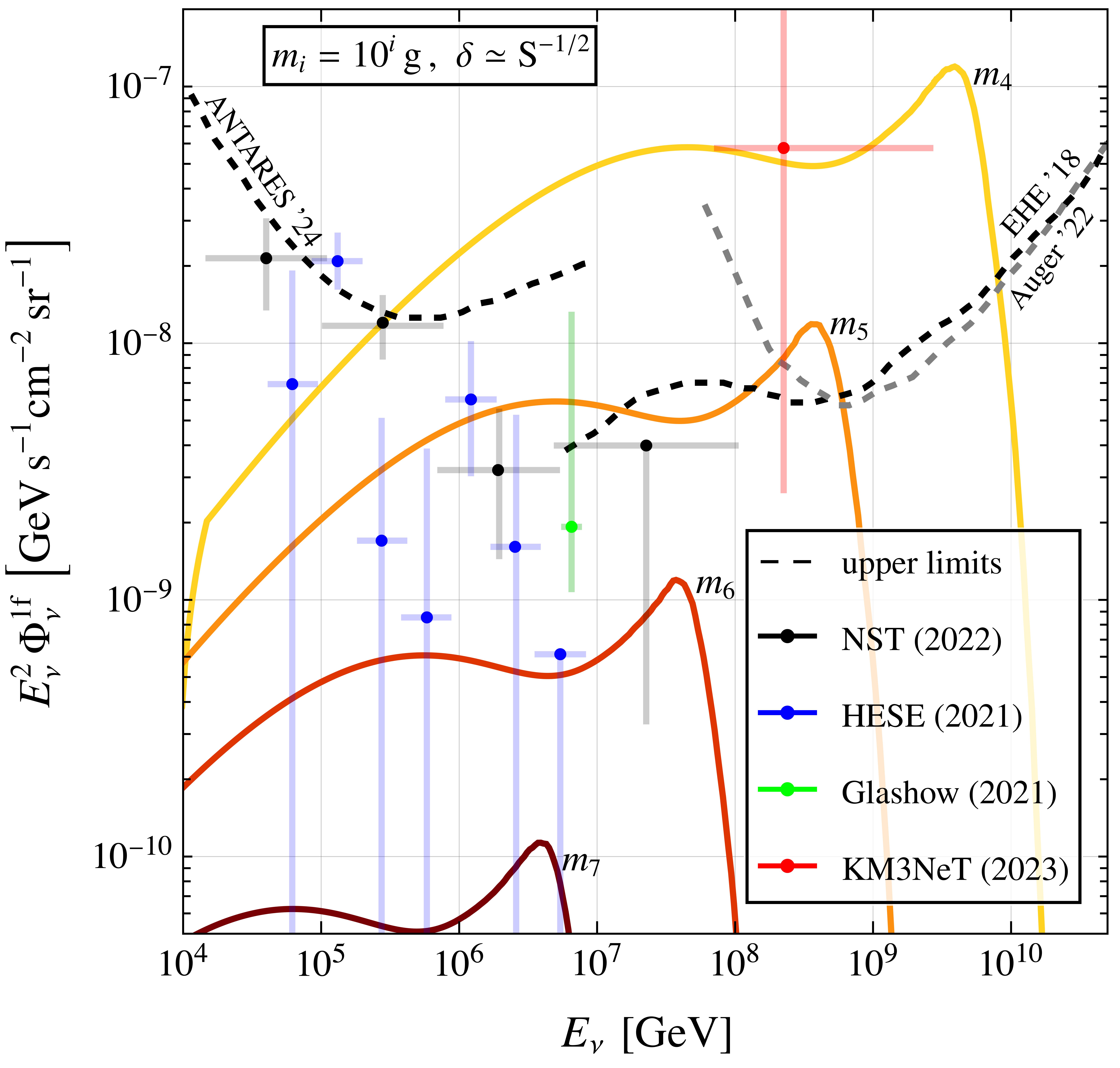}
    \caption{Per-flavor neutrino flux for different monochromatic PBH mass distribution with $f_{\rm PBH}=1$. Dashed lines denote existing upper bounds from IceCube High-Energy Starting Event (HESE)~\cite{IceCube:2018fhm} (black), Pierre Auger~\cite{PierreAuger:2023pjg} (gray) and ANTARES~\cite{ANTARES:2024ihw} (black). The black (blue) points represent the IceCube detection from Northern Sky Tracks, NST~\cite{Abbasi:2021qfz}  (HESE~\cite{IceCube:2020wum}).  
    The green cross shows the Glashow resonance measured by IceCube~\cite{IceCube:2021rpz}. Similarly, the high-energy neutrino KM3-230213A~\cite{KM3NeT:2025npi} is shown in red.}     \label{fig:neutrinoflux}
\end{figure}

     Numerous analytic arguments as well as 
     numerical simulations of the prototype systems~\cite{Dvali:2018xpy,Dvali:2018ytn,Dvali:2020wft,Dvali:2021tez,Dvali:2021bsy,Dvali:2023qlk,
     Alexandre:2024nuo,Dvali:2024hsb} indicate the universality of the phenomenon, which 
     must be exhibited by black holes. 
   Various  consistency and analyticity requirements show that  the lifetime of a burdened black hole is expected to be prolonged by the extra powers of the initial entropy~\cite{Dvali:2020wft,Alexandre:2024nuo,Dvali:2024hsb}, 
\begin{equation} \label{tMB}
	 \tau   \simeq \S^{1+k}\, r_g \;,
\end{equation}
   where $k$ is a positive integer.  
          
  The previous studies made it clear that the MB effect sets in 
latest by the time the system loses half of its initial mass. 
However, in reality the onset of MB 
can take place much earlier. 
 Not surprisingly, when applied to black holes, 
 the MB effect has dramatic consequences for 
 PBHs. In particular, it opens up a new window 
 of light PBH dark matter in the mass range below $10^{15}\,\rm g$~\cite{Dvali:2020wft, Alexandre:2024nuo,Thoss:2024hsr,Dvali:2024hsb}, which under standard 
 semi-classical treatment would be assumed to be long gone.
  It also softens the existing constraints on heavier black holes. 

 Of course, since the loss of the mass required for reaching the MB phase
 is gradual, there exists a smooth transition period.  
However, phenomenological studies~\cite{Dvali:2021byy,Franciolini:2023osw, Alexandre:2024nuo, Thoss:2024hsr,Dvali:2024hsb,Balaji:2024hpu,Haque:2024eyh,Barman:2024iht,Bhaumik:2024qzd,Barman:2024ufm,Kohri:2024qpd,Borah:2024bcr,Chianese:2024rsn,Zantedeschi:2024ram,Barker:2024mpz,Loc:2024qbz,Basumatary:2024uwo,Federico:2024fyt,Athron:2024fcj,Barman:2024kfj,Bandyopadhyay:2025ast,Calabrese:2025sfh,Boccia:2025hpm,Liu:2025vpz} have so far made the 
simplified approximation of a sudden onset of MB, \ie that the evaporation rate changes sharply from the semi-classical value to MB-suppressed rate leading to the prolonged lifetime \eqref{tMB}.

 In this paper, we shall go beyond this assumption and account for 
 a smooth transition to the MB state.  We demonstrate
that the transition phase has a crucial effect on observational
constraints.  In particular, for a prolonged transition 
the evaporation will persist at relevant levels until today. 
Since this can lead to constraints on the abundance of PBHs, our paper significantly updates the previous 
 studies of astrophysical fluxes of MBPBHs~\cite{Thoss:2024hsr,Chianese:2024rsn,Zantedeschi:2024ram,Liu:2025vpz}.
 Strikingly, some of already observed neutrino events might be explained by currently-transitioning MBPBHs, as shown in \fig \ref{fig:neutrinoflux}.

\paragraph*{Essence of memory burden}
For completeness of the logic flow, 
we shall briefly recap the key features of the MB effect and its manifestations in black holes. 
  For detailed derivations and analysis the reader is referred to the series of original articles~\cite{Dvali:2018xpy,Dvali:2018ytn,Dvali:2020wft,Dvali:2024hsb}. 

We start from the free Hamiltonian of $K$ memory modes
\begin{equation} \label{HamiltonianFree}
	\hat{H}_{\text{free}} = \epsilon_K \sum_{k=1}^K  \hat{n}_k   \,,
\end{equation}
where $[\hat{a}_j,\hat{a}_k^{\dagger}] = \delta_{jk}$, $[\hat{a}_j,\hat{a}_k]  =   [\hat{a}_j^{\dagger},\hat{a}_k^{\dagger}] =0$ correspond to usual bosonic commutation relations,  $\epsilon_K$ is the free energy gap, and $\hat{n}_k\equiv \hat{a}_k^{\dagger} \hat{a}_k$ denote number operators with expectation values $n_k\equiv \braket{\hat{n}_k}$. The bosonic nature 
of the memory modes is not important for our analysis.  
Even if we were to restrict ourselves to fermionic qubits, $n_k = 0, 1$, the Hilbert space of the theory would have  exponentially large dimensionality of $2^K$ with basis of number eigenstates,  $\ket{n_1, \ldots, n_K}$.
Each sequence of the occupation numbers, 
represents an information pattern, or a  ``memory pattern".

   An universal feature of systems of high efficiency 
   of information storage is that the memory states of high microstate degeneracy are reached
   for special values of certain order parameter.  
    This order parameter is formed by a high occupation number of a low energy mode, called a master mode. 
  The shared feature of high memory capacity 
  systems is that, due to the attractive interaction between 
  master and memory modes, for certain critical  occupation number of the master mode, the memory modes become gapless. 

   The essence of the MB phenomenon~\cite{Dvali:2018xpy,Dvali:2018ytn,Dvali:2020wft,Dvali:2024hsb} can be  explained by considering the 
   following prototype Hamiltonian 
\begin{equation} \label{Hamiltonian}
	\hat{H} = \epsilon_0 \hat{n}_0 + \epsilon_K \left(1-\frac{\hat{n}_0}{N_c}\right)^p \sum_{k =1}^K  \hat{n}_k  \,.
\end{equation}
 This Hamiltonian was originally proposed in~\cite{Dvali:2017nis, Dvali:2018vvx,Dvali:2018tqi} for describing the universal mechanism of ``assisted gaplessness"  
 of the memory modes in systems of high information storage capacity, such as black holes, without considering the MB effect. The MB phenomenon was discovered later 
 in ~\cite{Dvali:2018xpy,Dvali:2018ytn,Dvali:2020wft,Dvali:2024hsb} and studied in subsequent papers.
In \eq \eqref{Hamiltonian}, $\hat{n}_0$ is the number operator of the master mode, the gap of which is $\epsilon_0$. Moreover, the strength of the attractive interaction is set by $N_c$ and the exponent $p$ is chosen to be an integer.
The crucial role of the $\hat{n}_0$-mode is to decrease the effective energy gap of the $\hat{n}_k$-modes:
\begin{equation} \label{effectiveGap}
	\mathcal{E}_K = \left(1-\frac{n_0}{N_c} \right)^p\epsilon_K \,.
\end{equation}
Indeed, for the critical occupation $n_0 = N_c$, all $\hat{n}_k$-modes become gapless, $\mathcal{E}_k=0$ -- the master mode assists the memory modes in becoming gapless~\cite{Dvali:2018tqi}. Therefore,  all states 
\begin{equation}
	\ket{\underbrace{N_c}_{n_0}, n_1, \ldots, n_K}
\end{equation}
become degenerate in energy for arbitrary values of $n_1$, \ldots, $n_K$. This exponential number of states leads to efficient information storage measured by a microstate entropy  $\sim K$.

   For any nontrivial memory pattern the  critical point 
of gaplessness, $n_0 = N_c$,
represents a local minimum of energy. 
 Any change in the occupation number of the master mode
 requires a climb uphill the energy barrier which creates the resistance of the system against the loss of the master mode. 
This is the essence of the MB effect \cite{Dvali:2018xpy,Dvali:2018ytn,Dvali:2020wft,Dvali:2024hsb}. Quantitatively, we can understand this backreaction in terms of an effective contribution to the gap of the master mode
\begin{equation} \label{mu}
	\mu = N_m \left|\frac{\partial  \mathcal{E}_K}{\partial n_0}\right| = \frac{p N_m}{N_c} \left(1-\frac{n_0}{N_c} \right)^{p-1} \epsilon_K \,,
\end{equation}
where $N_m \equiv \sum_{k =1}^K n_k$ is the total occupation in the memory sector.  For $p\geq 2$, we see that  for the initial occupation, $n_0=N_c$,  MB is zero, $\mu =0$. However,
it increases rapidly once $n_0$ departs from its critical value $N_c$. 

Correspondingly, the loss of the master mode is effectively blocked  the latest when $\mu$ becomes of the order of the free gap $\epsilon_0$. This stage signals the full transition to the MB phase.  
The corresponding critical relative loss of particles is given by \cite{Dvali:2018xpy,Dvali:2018ytn,Dvali:2020wft,Dvali:2024hsb}
\begin{equation} \label{NCrit}
q\equiv	\frac{\left(N_c-n_0\right)_{\text{crit}}}{N_c} = \left(\frac{\epsilon_0 N_c}{p \epsilon_K N_m}\right)^{1/(p-1)} \;.
\end{equation}

\paragraph*{Application to black holes}

 The studies performed in the previous papers~\cite{Dvali:2018xpy,Dvali:2018ytn,Dvali:2020wft,Dvali:2024hsb} show that  thanks to the universality of the MB effect the Hamiltonian \eqref{Hamiltonian} captures 
 its essence in a large variety of systems of enhanced 
 information storage, including black holes. 
  The specific differences are fully determined 
  by parameters such as $\epsilon_0, \epsilon_K, N_c, K$ and the critical exponent $p$.
   For a black hole of mass $M$, the dictionary is~\cite{Dvali:2018xpy,Dvali:2018ytn,Dvali:2020wft,Dvali:2024hsb}
\begin{equation} \label{parametersBH}
	\epsilon_0=r_g^{-1} \,, \quad N_c=K=\S\,,\quad N_m=\S/2\,, \quad \epsilon_k = \sqrt{\S} r_g^{-1} \;.
\end{equation}
  The only remaining quantity, which  we shall treat 
  as a free input parameter, is the critical exponent 
  $p$. 
  
Plugging this mapping into the critical value \eqref{NCrit} of lost particles, we obtain
\begin{equation} \label{q}
q \simeq \left(p^2 \S\right)^{-1/(2(p-1))} \;.
\end{equation}
While MB essentially sets in immediately for $p=1$, we have
\begin{equation} \label{p2}
	p=2 \qquad \Rightarrow \qquad q=\frac{\Delta M_{\text{crit}}}{M} = \frac{1}{\sqrt{\S}} \;.
\end{equation}
This possibility of a breakdown of the semi-classical description after the timescale $\sqrt{\S}\, r_g$ receives independent motivation from criticality of a black hole state~\cite{Dvali:2011aa,Dvali:2012rt,Dvali:2012en,Dvali:2012wq,Dvali:2013vxa}, which also points to the scaling \eqref{p2},  $\Delta M_{\text{crit}}/M = 1/\sqrt{\S}$ \cite{Dvali:2015wca,Michel:2023ydf}. Of course, $p$ can also be much larger:
\begin{equation}
		p\gtrsim \ln \S \qquad \Rightarrow \qquad q=\frac{\Delta M_{\text{crit}}}{M} \sim O(1) \;,
\end{equation}
enabling the semi-classical description to remain valid up until half-decay, 
$q=1/2$. This would also coincide with so-called Page's time~\cite{Page:1993wv}, although 
in the present case the physical meaning is drastically different. 

 	\paragraph*{Transition to MB}
	As key novelty of this work, we shall now present different estimates for the transition to MB.
Physically, the transition period defines an intermediate state 
  in which while master mode is affected, the interaction rates of the memory modes are still highly suppressed due to their small gaps.  
    Correspondingly, the system cannot get rid of the information load. 
    This back reaction slows down the particle emission to the rate that  we estimate as,
    \begin{equation} \label{GammaDeltaN}
	\Gamma =  \left(\frac{1}{\S}\right)^{\Delta N} \ \Gamma_{\rm sc} \;,
\end{equation}
 where     
	\begin{equation} \label{DeltaN}
		\Delta N = \mu\, r_g  = \frac{p \sqrt{\S}}{2} \left(\frac{M_0 - M(t)}{M_0} \right)^{p-1}\;,
	\end{equation}
and we used \eq \eqref{mu} together with the black hole correspondence \eqref{parametersBH}.
Notice that $\Delta N$ measures the gap of the master mode in units 
of the black hole radius. 
 Since $\Delta N$ is a function of $M(t)$ by \eq \eqref{DeltaN}, $M(t)$ is determined by
\begin{equation} \label{multiparticleEq}
		\frac{\diff M(t)}{\diff t} = - r_g^{-1}\,\Gamma \;,
\end{equation}
where we note that $\Gamma\simeq r_g^{-1}$ would reproduce the semi-classical result.

 It is very important to understand that, in contrast to what is assumed in 
 extrapolations of the semi-classical treatment, 
 due to the MB effect the evaporation of a black hole cannot be self-similar. Correspondingly, it makes not much sense to 
 assume that the quantities  $\Gamma_{\rm sc}$  and  $r_g$ 
 track the mass of a black hole according to semi-classical dependence  on the black hole mass. Instead, in our treatment $\Gamma_{\rm sc}$ 
 and $r_g$ are understood as the time-independent parameters given 
 by the semi-classical rate and the radius at the initial time. 
Making them time-dependent  according to the semi-classical mass-dependence, although only creating a minor error, would be incorrect.

\paragraph{Width of transition}
 Expanding \eq \eqref{DeltaN} around the onset of MB, we estimate 
\begin{equation} \label{GammaDeltaNScaling}
	\Gamma \simeq \exp\left(-\frac{(1-q) M_0 - M(t)}{\delta M_0}\right) \ \Gamma_{\rm sc} \;,
\end{equation}
where we defined
\begin{equation} \label{delta}
	\delta \equiv \frac{q}{(p-1)\ln \S} \;.
\end{equation}
For $p=2$, this gives $\delta \simeq 0.02\, q$ while for $p\simeq \ln \S $ we would get $\delta \simeq 4 \times 10^{-4} q$. Eq.\@\xspace \eqref{GammaDeltaNScaling} shows that $q$ determines the critical mass loss at which MB sets in, while $\delta$ determines the width of the transition, \ie the mass loss relative to $M_0$ during the crossover period.

In order to confirm this estimate, we shall employ a second approach for determining $\delta$.
We can say that MB is fully reached when
\begin{equation}\label{eq:deltancritmurg}
	r_g\, \mu\big(M(t)=M_0(1-q)\big) = \Delta N_{\rm crit} \;,
\end{equation}
where $\Delta N_{\rm crit}$ is a small integer (\eg $\Delta N_{\rm crit}\approx 2$ from setting $(1/\S)^{\Delta N_{\rm crit}} = (1/\S)^{k}$; \cf \eq \eqref{GammaDeltaN}). Therefore, already for a small value of $\Delta N$, the rate-suppression is comparable to the one expected in the full MB phase, leading to the lifetime \eqref{tMB}.

From \eqref{eq:deltancritmurg} we get
\begin{equation}
	q =\left(\frac{p \sqrt{\S}}{\Delta N_{\rm crit}}\right)^{-1/(p-1)} \;.
\end{equation}
Moreover, the beginning of the onset of MB corresponds to
\begin{equation}
	\mu\big(M(t)=M_0(1-q+\delta)\big)= r_g^{-1} \;,
\end{equation}
which gives
\begin{equation}\label{eq:deltawidth}
	\delta = q - \left(p \sqrt{\S}\right)^{-1/(p-1)} = q \left(1- \Delta N_{\rm crit}^{-1/(p-1)}\right) \;.
\end{equation}
As an example, $ \Delta N_{\rm crit} =2$ and $p=2$ give $\delta =0.3 q$ while $p=50$ would imply $\delta \approx 0.014 q$. Within order of magnitude these estimates are consistent with our previous approach. We observe that $\delta$ has to be smaller than $q$ but the separation of the scales is not large. In other words, the main reason for obtaining a very small $\delta$ is a correspondingly small $q$, \ie an early onset of MB.

One can certainly ask how general is the  prototype Hamiltonian \eqref{Hamiltonian} for capturing the main aspects of the transition period
to MB. One aspect of reliability is that MB effect is a type of a critical phenomenon. It is typical that near criticality the behavior of the systems 
is well captured by leading exponents. Therefore, approximation of 
the gap with a monomial, with the critical exponent $p$, is expected 
to capture the order of magnitude estimates reliably. 
Of course, this said, one has to be open-minded in exploring 
alternative gap functions for phenomenological studies. 

Taking into account the rate \eqref{GammaDeltaNScaling}, \eq \eqref{multiparticleEq} can be solved explicitly
\begin{equation} \label{sol}
(1-q) M_0 -M(t) = \delta M_0 \ln\left(\frac{t}{\delta \tau_{\text{SC}}} + 1 - \frac{q}{\delta} \right) \;,
\end{equation}
to show that
\begin{equation}\label{dmdtanalyticscaling}
	\frac{\diff M(t)}{ \diff t} = -r_g^{-1}\,\Gamma_{\rm sc}\frac{\delta}{\left(\frac{t}{\tau_{\text{SC}}}-(q-\delta)\right)} \;,
\end{equation}
where we expressed the result in terms of the semiclassical would-be lifetime $\tau_{\text{SC}}$.  
For large $t$, we can approximate
\begin{equation} \label{scaling1/t}
\frac{\diff M(t)}{\diff t} \simeq -r_g^{-1}\,\Gamma_{\rm sc} \frac{\delta \, \tau_{\text{SC}} }{t} \;.
\end{equation} 
Therefore, on timescales larger than $\tau_{\rm SC}$, the energy injection is independent of $q$.
We show the solution to \eq \eqref{multiparticleEq} for different values of $p$ and compare it to the analytic scaling \eqref{dmdtanalyticscaling} varying $q$ and $\delta$ in \fig \ref{fig:comparison}, where we assumed $k\gtrsim 2$. Further values are shown in the appendix in \figs \ref{fig:multiparticle} and \ref{fig:analytic}.

\paragraph*{Comparison with observations}
\begin{figure}
	\centering
	\includegraphics[width= 0.9\linewidth]{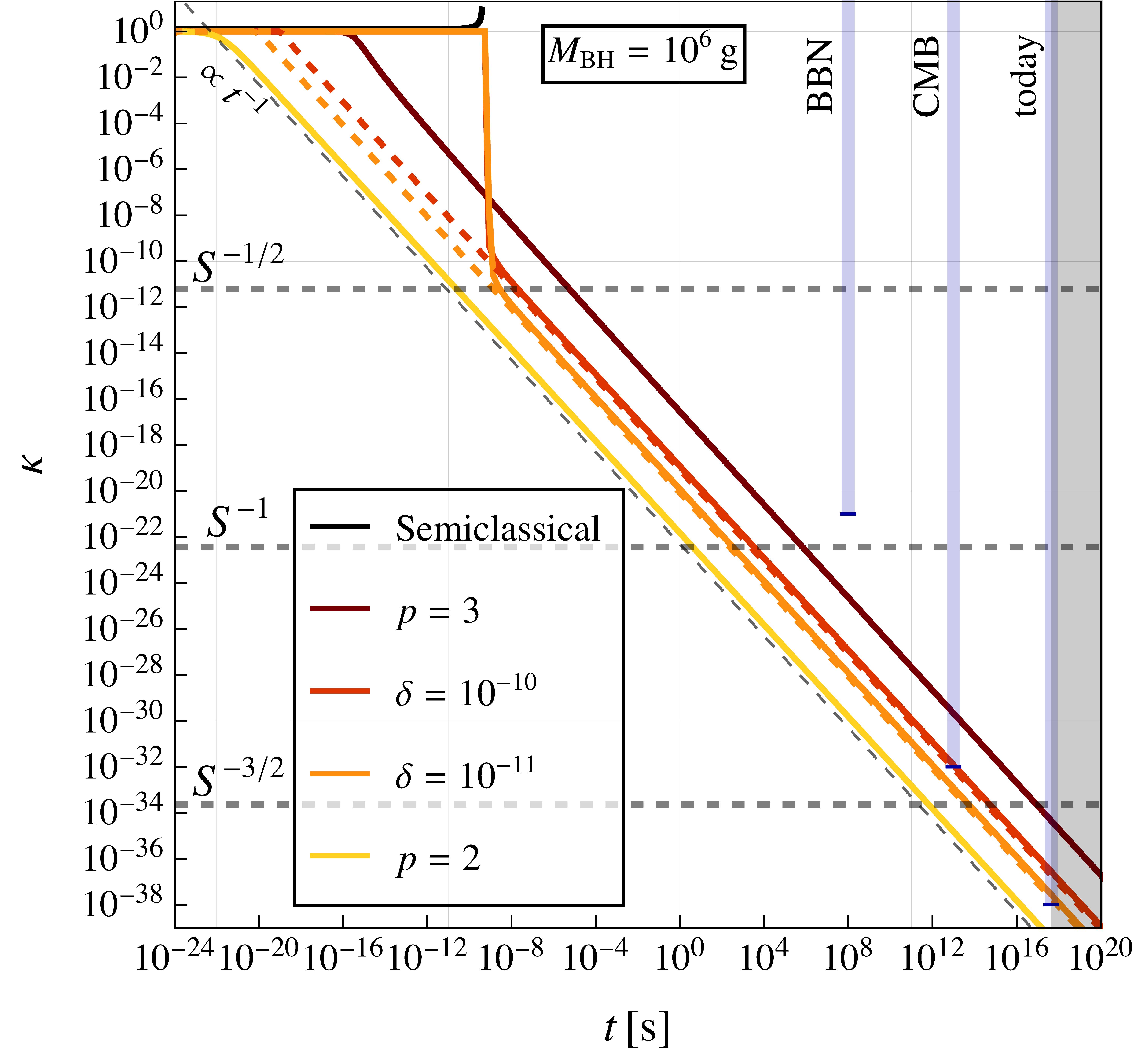}
	\caption{Solution of \eq \eqref{multiparticleEq} for $p=2,3$ as well as analytic solution \eqref{sol} for $\delta= 10^{-10},10^{-11}$ and $q=1/2$ are shown. Dashed-lines correspond to the same choice of $\delta$ and $q=\delta$. It is clear that the choice of $q$ is irrelevant for large times. 
    Moreover, the constraints estimated in \eqs \eqref{constraintBBN}, \eqref{constraintCMB} and \eqref{constraintToday} are schematically displayed.}
	\label{fig:comparison}
\end{figure}

The relevant quantity for comparison with observational constraints is
\begin{equation}
\label{eq:kappadefinition}
\kappa \equiv \frac{\diff M}{\diff t}\Big/ \frac{\diff M(0)}{\diff t} \;, 
\end{equation}
\ie the relative suppression of the evaporation rate at a given time \wrt the initial one. 
The constraints from BBN, CMB and from today's Universe obviously depend on the relative emission energy of the black hole, and therefore, on its mass. Although even PBHs heavier than $10^{10}\,\rm g$ could constitute all of the dark matter for sufficiently small $q$~\cite{Thoss:2024hsr}, we shall focus on the lighter window which is viable also for $q\sim \mathcal{O}(1)$  due to its phenomenological relevance.

As discussed in the appendix, the resulting constraints are estimated as
\begin{align}
\text{BBN:}\quad	\kappa &\lesssim  10^{-21} \left(\frac{M_0}{10^6\,\text{g}}\right)^3 \quad \text{at} \quad t\simeq 10^8\, \text{s} \;, \label{constraintBBN}\\
\text{CMB:}\quad	\kappa &\lesssim  10^{-32} \left(\frac{M_0}{10^6\,\text{g}}\right)^3 \quad \text{at} \quad t\simeq 10^{13}\, \text{s} \;,\label{constraintCMB}\\
\text{today:} \quad	\kappa &\lesssim 10^{-38} \left(\frac{M_0}{10^6\,\text{g}}\right)^3 \quad \text{at} \quad t\simeq  10^{17}\, \text{s} \;.\label{constraintToday}
\end{align}
From the scaling \eqref{scaling1/t} derived before, we see that
\begin{equation}
\label{eq:simplifiedscalingdmdt}
	\kappa = \delta \frac{10^{-10}\, \text{s}}{t} \left(\frac{M_0}{10^6\,\text{g}}\right)^3 \;.
\end{equation}
The discussion in appendix shows that the strongest constraint comes from today's Universe (it is actually two orders of magnitude stronger than \eg CMB constraints), which yields
\begin{equation}\label{eq:deltatoday}
	\delta \lesssim 10^{-38} \frac{10^{17}\, \text{s}}{10^{-10}\, \text{s}} \lesssim 10^{-11} \;,
\end{equation}
for $f_{\rm PBH}=1$. 

We confirm this result by plotting the bounds \eqref{constraintBBN}, \eqref{constraintCMB} and \eqref{constraintToday} together with the previously derived forms of the transition in \fig \ref{fig:comparison}. Evidently, the constraint \eqref{eq:deltatoday} does not explicitly depend on the black hole mass (see also discussion in appendix), but $\delta$ generically shrinks with larger $M_0$ and so \eq \eqref{eq:deltatoday} leads to a lower bound on the mass of PBHs as dark matter.

Quantitatively, we recall that $\delta \lesssim q$ (\cf\eq \eqref{delta}). For the well-motivated case $p=2$ (in terms of \eq \eqref{multiparticleEq}), we have $q=1/\sqrt{\S}$ and since $\S \gtrsim 10^{22}$ for $M_0\gtrsim 10^6\,\text{g}$, we get the following window for small PBHs as dark matter:
\begin{eqnarray} \label{eq:earlyBurdenWindow}
M_0 \gtrsim 10^6\,\text{g}\;, \quad	p=2 \;, \quad q = \frac{1}{\sqrt{\S}}\;, \quad  \delta \approx 0.1 q \;,
\end{eqnarray} 
corresponding to an early onset of MB after losing only a fraction $1/\sqrt{\S}$ of the initial mass.
The bound $M_0 \gtrsim 10^6\,\text{g}$ is very similar the one proposed in~\cite{Alexandre:2024nuo,Thoss:2024hsr} with the assumption of a sudden onset of MB. Larger values of $p$ also allow for PBHs as dark matter, albeit with larger masses. 

\paragraph*{Transition to MB as neutrino source}

We shall now complement our previous statements with an actual calculation of the neutrino flux expected today. To do so, we adopt the scaling in \eqref{eq:simplifiedscalingdmdt} for the energy-loss of the PBHs during their cosmological evolution. We assume that the PBHs is still in the transition to MB, compatible with a sufficiently large $k$ in \eqref{tMB}. Arbitrarily large values of $k$ can be probed contrary to the case of an approximate instantaneous transition~\cite{Thoss:2024hsr,Chianese:2024rsn,Liu:2025vpz}.

The resulting sum of the galactic and extra-galactic per-neutrino flavor-flux is shown in Fig.~\ref{fig:neutrinoflux} for different monochromatic PBH mass distributions with $f_{\rm PBH}=1$. 
Dashed lines correspond to existing upper bounds, while crosses show the existing neutrino detections, the details of which are reported in the Figure caption. 

In \fig \ref{fig:neutrinoflux}, we have chosen $\delta \simeq \,S^{-1/2}$ compatible with the analytic result \eqref{eq:earlyBurdenWindow} characteristic of the early onset of MB (different values of $\delta$ are allowed in the most general phenomenological parametrization \eqref{GammaDeltaNScaling}). Obviously, only the numerical value of $\delta$ matters, and we observe no constraints as long as $\delta \lesssim 10^{-11}$ (the equality corresponds to the case of $10^6\,$g in the figure), in line with the previous analytic discussion, \cf \eq \eqref{eq:deltatoday}. 

The transition to MB has a very peculiar behavior as a function of redshift -- in particular, the injected energy per physical volume scales approximately as $\propto (1+z)^3 \,t^{-1}$, which implies a larger cosmological signal as compared to a decaying dark matter.\footnote{Ref.~\cite{Zantedeschi:2024ram} pointed out that another source of astrophysical highly energetic particles -- leading to fluxes comparable to present-day observations -- is due to the merger of MBPBHs in the late Universe, which create ``young" black holes (see the discussion in~\cite{Dvali:2023qlk} on the merger dynamics) resuming their Hawking evaporation.}
Characterization of the counterpart signal in gamma and cosmic rays requires further investigation.

\paragraph*{Conclusion}

 In the present paper we have taken into account 
 the non-instantaneous nature of the transition of 
 PBHs to the MB phase. 
 The gradual transition has a crucial impact 
 on the observational bounds. In particular, for light PBHs that are still transitioning to MB, what matters is the slope of the transition rather than the parameter $k$ in the final rate (see \eq \eqref{tMB}).

 Most notably, it leads to new potentially detectable signals from  small PBHs that are currently transitioning 
 to the MB phase. 
 In particular, we point out that such transient 
 PBHs can be source of high energy neutrinos  
 within the reach of present day experiments such as IceCube and KM3NeT.

\begin{acknowledgments}
	
	\paragraph*{Acknowledgments}  M.Z.~acknowledges useful ongoing discussion on astrophysical high-energy particles with Daniele Gaggero, Giulio Marino and, in particular, Paolo Panci. 
     G.D.~is grateful to Elisa Resconi for discussions about
 IceCube and related topics. 
 
The work of G.D. was supported in part by the Humboldt Foundation under Humboldt Professorship Award, by the European Research Council Gravities Horizon Grant AO number: 850 173-6,
by the Deutsche Forschungsgemeinschaft (DFG, German Research Foundation) under Germany's Excellence Strategy - EXC-2111 - 390814868, and Germany's Excellence Strategy under Excellence Cluster Origins. The work of S.Z.~was supported by the European Research Council Gravites Horizon Grant AO number: 850 173-6. 

\paragraph*{Disclaimer} Funded by the European Union. Views
and opinions expressed are however those of the authors
only and do not necessarily reflect those of the European Union or European Research Council. Neither the
European Union nor the granting authority can be held
responsible for them.	
\end{acknowledgments}

\paragraph{Note Added} We are aware of the work~\cite{monte:2025} which also deals with BBN and CMB constraints from the transition to MB. While the analysis of~\cite{monte:2025} is consistent with ours, the conclusions are different. M.Z.~and S.Z.~acknowledge collaboration on that project.

\appendix
\section{Additional figures}
In this Section we show the evolution of the suppression factor $\kappa$ defined in \eqref{eq:kappadefinition} for the different differential equations discussed in the main text.

\begin{figure}
	\centering
	\includegraphics[width= 0.9\linewidth]{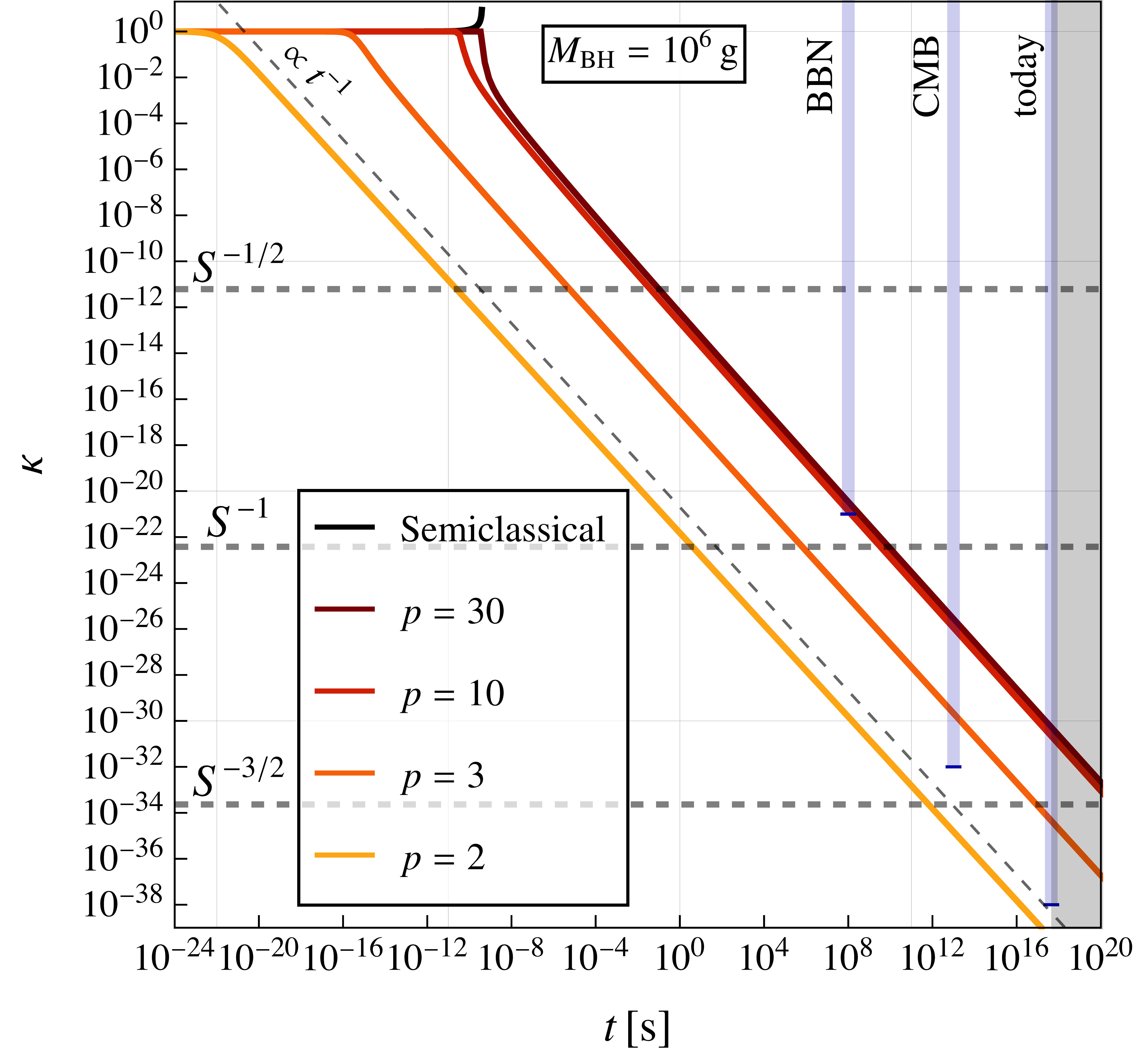}
	\caption{Solution of \eq \eqref{multiparticleEq} obtained adopting \eqref{GammaDeltaN} for different values of $p$.}
	\label{fig:multiparticle}
\end{figure}

Fig.~\ref{fig:multiparticle} displays the solution of \eq \eqref{multiparticleEq} starting from \eqref{GammaDeltaN} for different values of $p$ characterizing the prototype Hamiltonian~\eqref{Hamiltonian}. The timescale of onset of MB for different $p$ values is consistent with the analytic estimate we obtained for \eqref{q}. Similarly, the width of the transition $\delta$ is also approximately given by \eqref{eq:deltawidth}. For large enough $p$ the asymptotic behavior of the curve becomes similar as it can be verified by direct comparison of the case $p=10$ and $p=30$.

\begin{figure}
	\centering
	\includegraphics[width= 0.9\linewidth]{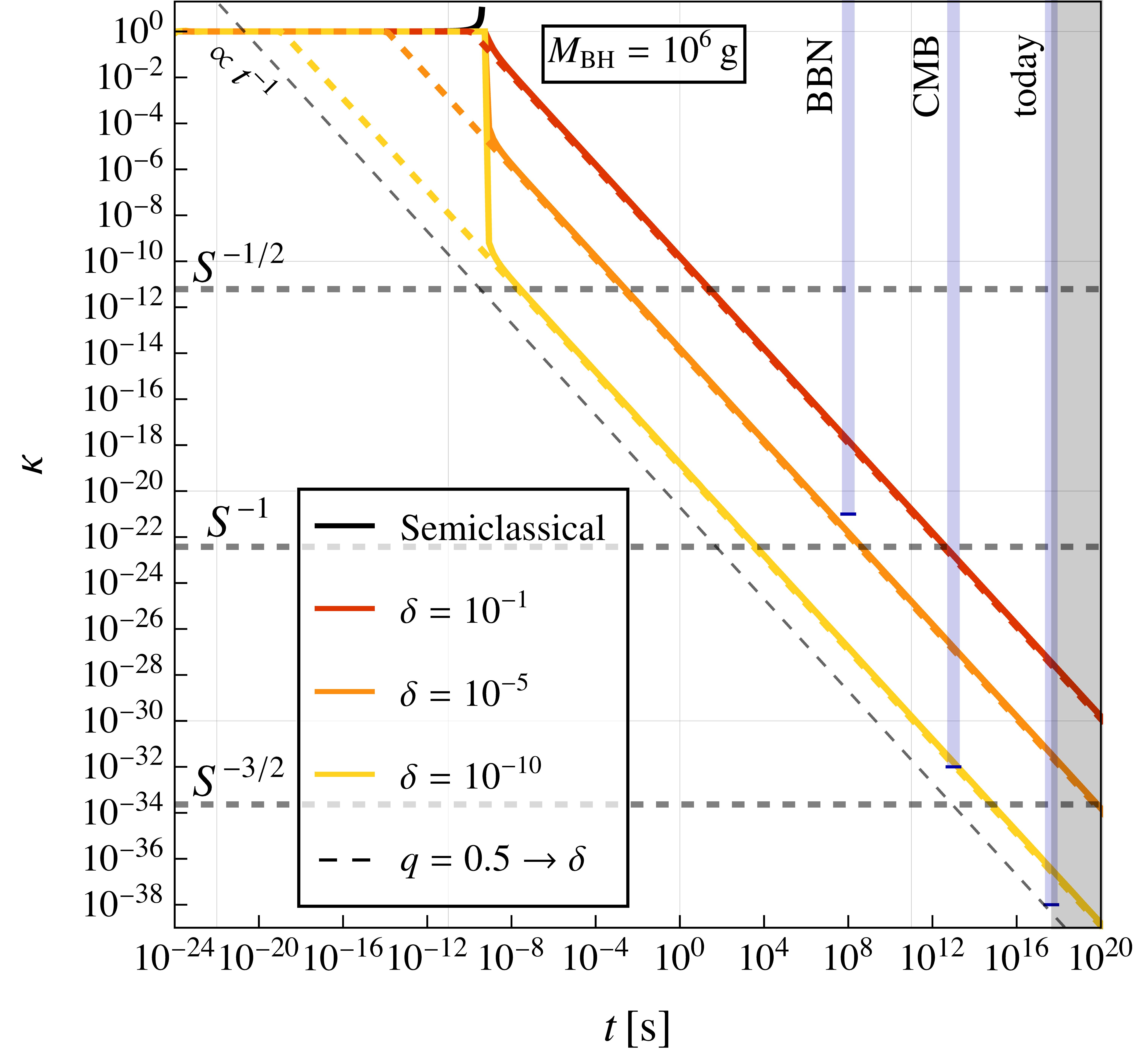}
	\caption{Solution of \eq \eqref{multiparticleEq} obtained adopting the analytic approximation \eqref{GammaDeltaNScaling} for different $\delta$. Here continuous (dashed) lines correspond to $q=0.5$ ($\delta$).}
	\label{fig:analytic}
\end{figure}

Fig.~\ref{fig:analytic} shows the general solution of \eq \eqref{multiparticleEq} taking into account approximation \eqref{GammaDeltaNScaling} for different values of $\delta$ and $q$. Different colors stands for different $\delta$ choices, while continuous (dashed) lines correspond to $q = 1/2$ ($q=\delta$) respectively. It is clear from the Figure that the asymptotic behavior is uniquely determined by the choice of $\delta$. In fact, the independence of the scaling from the choice of $q$ can also be understood from \eqref{scaling1/t}.

It is evident that regardless of the parametrization, both cases are well-described by $\kappa$ scaling as $1/t$. 


\section{Derivation of constraints}
As anticipated in the main text, the slow onset of MB might affect the Universe at both  early and late times. 

In order to get a feeling for such constraints, we shall adapt the approach of~\cite{Keith:2020jww}, and perform a comparison between the energy injection rate of a decaying dark matter particle $\chi$ with the one of the evaporating BHs approaching the MB phase. 
The former is given by
\begin{equation}
\label{eq:eninjdm}
    \left(\frac{{\rm d}E}{{\rm d}V \, {\rm d}t}\right)_{\rm DM} \simeq \frac{f_{\chi}}{\tau_{\chi}}\rho_{\rm c}\, (1+z)^3\,e^{-t/\tau_{\chi}}
\end{equation}
where in obvious notation, $f_{\chi} = \rho_\chi /\rho_{\rm DM}$ is the fraction between the energy density of the non-relativistic particle $\chi$ and the dark matter, $\tau_{\chi}$ its lifetime and $\rho_{\rm c}$ the critical energy density. 

For evaporating BHs, we get instead
\begin{equation}
\label{eq:eninjpbhs}
    \left(\frac{{\rm d}E}{{\rm d}V \, {\rm d}t}\right)_{\rm PBH} =\frac{{\rm d}M_{\rm BH}}{{\rm d}t}f_{\rm PBH}\, \frac{\Omega_{\rm DM}}{M_{\rm BH}}\, \rho_c(1+z)^3\,.
\end{equation}
 Requiring that the energy injection from PBHs does not  overshoot the one from decaying dark matter, and evaluating the expression at the time of particle decay, $t=\tau_{\chi}$, we arrive at
 \begin{equation}
 \begin{split}
     \label{eq:fpbhbound}
     f_{\rm PBH} \lesssim \frac{f_{\chi}\, M_{\rm BH}\,e^{-1}}{\tau_{\chi} \frac{{\rm d}M_{\rm BH}(t=\tau_{\chi})}{{\rm d} t}  \Omega_{\rm DM}} \\\simeq 10^2\frac{f_{\chi}\, M_{\rm BH}^3 \,G^2}{\tau_{\chi}\, \kappa\,   \Omega_{\rm DM}}
     &\simeq  \frac{2\,f_{\chi}\, \tau_{\rm SC}}{\tau_{\chi}\, \kappa\,   \Omega_{\rm DM}} \,,
     \end{split}
 \end{equation}
where we appropriately introduce the necessary numerical factor in the semiclassical rate to account for the evaporation in all Standard Model degrees of freedom. Furthermore, $\tau_{\rm SC}$ denotes the corresponding would-be half-mass semiclassical evaporation time and $\kappa$ is defined in \eqref{eq:kappadefinition}.

 Notice that 
 \eq \eqref{eq:fpbhbound} can be rewritten in an extremely suggestive form, in terms of physical characteristics of the nature of the two dark matter types considered
 \begin{equation}
 \label{eq:fpbhcmb}
     \kappa\lesssim 8\frac{f_{\chi}}{f_{\rm PBH}}\frac{\tau_{\rm SC}}{\tau_{\chi}} \,,\qquad \text{at }t\simeq \tau_\chi \simeq t_{\rm bound}.
 \end{equation}
We can now estimate the bound on the factor $\kappa$

Constraints from the CMB on decaying dark matter can be found in Ref.~\cite{Acharya:2019uba} (see also~\cite{Padmanabhan:2005es, Slatyer:2009yq, Chluba:2011hw,Slatyer:2016qyl,Poulin:2016anj,Cirelli:2024ssz}), while the ones from BBN are given in~\cite{Kawasaki:2017bqm,Jedamzik:2009uy}. Bounds on CMB and BBN for semiclassical PBHs are given in~\cite{Carr:2009jm,Keith:2020jww,Carr:2020gox,Haque:2024eyh,Boccia:2024nly}. A potential caveat is that PBHs in the ultralight mass window emit at energies $\sim  1/r_g$ which are much higher than the masses of decaying dark matter adopted in the above analyses. We notice that in both cases, the constraints become almost mass independent around $10^{13}\,$s for CMB~\cite{Acharya:2019uba} and $10^{7}\,$s for BBN respectively (a bound with slightly smaller constraining power follows at $t=1\,{\rm s}$, see~\cite{Kawasaki:2017bqm}). We therefore conservatively read off the bounds around these two times.

Present-day constraints on decaying dark matter due to gamma rays, neutrino fluxes and cosmic rays lead to stronger bounds. In order to see this, we notice that the present constraints on dark matter with mass between $10^4\,\rm GeV - 10^{12}\,\rm GeV$ decaying into quarks~\cite{LHAASO:2022yxw,Das:2023wtk}, neutrinos~\cite{Kohri:2025bsn} or photons~\cite{LHAASO:2022yxw,Das:2023wtk} all require a similar lifetime, which is conservatively of order $\tau_{\chi}\gtrsim 10^{29}\,{\rm s}$. These decay channels, however, are precisely the ones in which the PBHs also decay into.

Under the scaling ansatz \eqref{eq:simplifiedscalingdmdt}, the constraints on $\kappa$ reads
\begin{equation}
\label{eq:constrfdelta}
    f_{\rm PBH}\, \delta \lesssim 8\,f_{\chi}.
\end{equation}
It follows that BBN, CMB and present day astrophysical highly energetic particles constraints $f_{\rm PBH}\,\delta\lesssim \{10^{-3}, 10^{-9}, 10^{-11}\}$ respectively. This explains the origin of \eqref{constraintBBN}, \eqref{constraintCMB} and \eqref{constraintToday} and \eqref{eq:deltatoday}, where we dropped $\mathcal{O}(1)$ numerical factors compatibly with the precision of our estimates. We notice that \eqref{eq:constrfdelta} leads to constraints similar to those of~\cite{monte:2025} for CMB and BBN.
As shown by the analysis of the neutrino flux, the present-day constraints happen to be properly estimated by \eqref{eq:constrfdelta}.

\section{Calculation of neutrino flux}

The total flux $\Phi_{\nu}$ is comprised of two components, a galactic and an extragalactic one
\begin{equation}
    \Phi_{\nu}\doteq  \frac{{\rm d} \Phi_{\rm gal}}{{\rm d}E\, {\rm d}\Omega} + \frac{{\rm d} \Phi_{\rm egal}}{{\rm d}E\, {\rm d}\Omega}\,.
\end{equation}
The galactic all-flavor flux of neutrinos is given by
\begin{equation}
\label{eq:gal}
    \frac{{\rm d} \Phi_{\rm gal}}{{\rm d}E\, {\rm d}\Omega} = \frac{f_{\rm PBH}\bar{\mathcal{J}}}{ M_{\rm BH}}\frac{{\rm d}^2N}{{\rm d }E\,{\rm d}t}\,,
\end{equation}
where the averaged factor $\bar{\mathcal{J}}\simeq 2 \times 10^{22}\,\rm GeV\,cm^{-2}$~\cite{Cirelli:2012ut,Cirelli:2010xx}. 

Also, the extragalactic flux is given by
\begin{equation}
\label{eq:egal}
    \frac{{\rm d} \Phi_{\rm egal}}{{\rm d}E\, {\rm d}\Omega} = \frac{n_{\rm PBH}}{4\pi}\int_{\tau_{\rm SC}}^{t_0} {{\rm d} t}\, (1+z) \frac{{\rm d}^2N}{{\rm d }E\,{\rm d}t}\left((1+z) E \right)\,,
\end{equation}
where the density of PBHs today is $n_{\rm PBH}\simeq 2 \times 10^{-33} {\rm cm}^{-3} f_{\rm PBH}\, (10^{3}{\rm g}/M)$. In (\ref{eq:gal},\ref{eq:egal}) the particle number energy rate is summed over all neutrino species.

 Notice that $\frac{{\rm d}^2N}{{\rm d }E\,{\rm d}t}$ is given by
\begin{equation}
\label{eq:particleratetime}
    \frac{{\rm d}^2N}{{\rm d }E\,{\rm d}t} \simeq \kappa  \frac{{\rm d}^2N (t=0)}{{\rm d }E\,{\rm d}t}\,,
\end{equation}
where $\kappa$ is the suppression factor entering the energy rate \eqref{eq:kappadefinition}. Therefore,  $\frac{{\rm d}^2N}{{\rm d }E\,{\rm d}t}$ is time-dependent due to the scaling \eqref{eq:simplifiedscalingdmdt}.  

To evaluate the initial particle rate in \eqref{eq:particleratetime}, we use 
the numerical code {\tt BlackHawk 2.2}~\cite{Arbey:2019mbc, Arbey:2021mbl} which includes the emission in all Standard Model degrees of freedom. We obtained the total (including secondary) neutrino emission chosing the HDMSpectra option~\cite{Bauer:2020jay}.

If today the dark matter is already in the MB phase, $\kappa = S^{-k}$, in accordance with \eqref{tMB}. Then, the resulting flux is comparable to existing astrophysical particle fluxes for certain range of masses~\cite{Thoss:2024hsr,Chianese:2024rsn} (e.g., for $k=2$ this is true for masses $\lesssim 10^6\,\rm g$). This is extremely sensitive to the value of $k$.

\bibliography{Refs}

\end{document}